\documentclass[12pt,aps,prb,floatfix,amsmath,amssymb,superscriptaddress]{revtex4}
\usepackage{graphicx,subfigure}

\usepackage{amsthm,amssymb,amsmath,comment}
\usepackage{graphicx}

\usepackage{hyperref}
\usepackage{subfigure}
\usepackage{color}

\newcommand{\abs}[1]{\left\lvert#1\right\rvert}

\begin{document}

\title{Numerical integration for ab initio many-electron self energy calculations within the GW approximation}

\author{Fang Liu}
\email{fliu@lsec.cc.ac.cn}
\affiliation{School of Statistics and Mathematics, Central University of
Finance and Economics, Beijing 100081, China.}
\author{Lin Lin}
\email{linlin@lbl.gov}
\affiliation{Computational Research Division, Lawrence Berkeley National
Laboratory, Berkeley, CA 94720, USA.}
\author{
Derek Vigil-Fowler}
\email{vigil@berkeley.edu}
\affiliation{Department of Physics, University of California, Berkeley, CA 94720, USA.}
\author{
Johannes Lischner}
\email{jlischner597@gmail.com}
\affiliation{Department of Physics, University of California, Berkeley, CA 94720, USA.}
\author{
Alexander F. Kemper}
\email{afkemper@lbl.gov}
\affiliation{Computational Research Division, Lawrence Berkeley National
Laboratory, Berkeley, CA 94720, USA.}
\author{
Sahar Sharifzadeh}
\email{ssharifzadeh@lbl.gov}
\affiliation{Molecular Foundry, Lawrence Berkeley National
Laboratory, Berkeley, CA 94720, USA.}
\author{Felipe Homrich da Jornada}
\email{jornada@berkeley.edu}
\affiliation{Department of Physics, University of California, Berkeley, CA 94720, USA.}
\author{
Jack Deslippe}
\email{jdeslippe@lbl.gov}
\affiliation{NERSC, Lawrence Berkeley National
Laboratory, Berkeley, CA 94720, USA.}
\author{
Chao Yang}
\email{cyang@lbl.gov}
\affiliation{Computational Research Division, Lawrence Berkeley National
Laboratory, Berkeley, CA 94720, USA.}
\author{
Jeffrey B. Neaton}
\email{jbneaton@lbl.gov}
\affiliation{Department of Physics, University of California, Berkeley, CA 94720, USA.}
\affiliation{Molecular Foundry, Lawrence Berkeley National
Laboratory, Berkeley, CA 94720, USA.}
\affiliation{Materials Sciences Division, Lawrence Berkeley
National Laboratory, Berkeley, CA 94720, USA.}
\author{
Steven G. Louie}
\email{sglouie@berkeley.edu}
\affiliation{Department of Physics, University of California, Berkeley, CA 94720, USA.}
\affiliation{Materials Sciences Division, Lawrence Berkeley
National Laboratory, Berkeley, CA 94720, USA.}

\begin{abstract}
We present a numerical integration scheme for evaluating the
convolution of a Green's function with a screened Coulomb potential
on the real axis in the GW approximation of the self energy.
Our scheme takes the zero broadening limit in Green's function
first, replaces the numerator of the integrand with a piecewise
polynomial approximation, and performs principal value integration
on subintervals analytically. We give the error bound of our
numerical integration scheme and show by numerical examples that
it is more reliable and accurate than the standard quadrature rules
such as the composite trapezoidal rule. We also discuss the benefit of using
different self energy expressions to perform the numerical convolution
at different frequencies.

\end{abstract}

%\pacs{}
%\end{frontmatter}

\maketitle

\section{Introduction}
The computational modeling and simulation of single-electron
excitations of molecules and solids including many-electron effects is important for interpreting
spectroscopy experiments and predicting excited-state properties of materials.
One way to calculate single-particle excitation energies or
quasiparticle energies is through Green's function
theory \cite{Hedin65,Hybertsen86,Aryasetiawan98,Aulbur00,Onida02}. In such a theory, the quasiparticle energies
$\varepsilon_j$'s and wavefunctions $\phi_j(r)$'s are obtained
by solving Dyson's equation \cite{Hybertsen86,Hedin70} (in atomic units $\hbar=1$, $e=1$,
and electron mass $m_e=1$)
\begin{equation}
\left[-\frac{1}{2} \nabla^2(r) + v_H(r) \right]\phi_j(r)
+\int dr' \Sigma(r,r';\varepsilon_j) \phi_j(r') = \varepsilon_j \phi_j(r),
\label{eq:dyson}
\end{equation}
where $v_H(r)$ is the Hartree potential and
$\Sigma(r,r';\varepsilon_j)$ is the energy (or frequency) dependent self energy
operator.

The main challenge in solving \eqref{eq:dyson} is the approximation
and evaluation of $\Sigma$. One widely used approximation is the $GW$
approximation in which $\Sigma$ is set to the product of
a single-particle Green's function $G$ and a screened Coulomb potential
$W$. In the frequency domain, the product is evaluated as
a convolution of the time-ordered Fourier transforms of these two functions.
Due to the presence of singularities in both functions, the numerical
convolution must be carried out with care.  This is particularly
important for molecules because the poles of $W$ for such systems
have more discrete features than those in solids. These poles must be
treated properly in the self energy calculation.

There are a number of ways to perform the energy dependent (or
full frequency) self energy calculation numerically. One way
is use analytic continuation method proposed in \cite{Daling91,
Jin99,Rojas95}, another is
to perform a contour deformation \cite{Kotani02,Lebegue03}. A more direct approach is
to perform the numerical integration on the real axis, which is
the approach we take in this paper. All of these approaches
have positive and negative aspects and each approach represents a different way to overcome
potential numerical difficulties introduced by the singularity
of the integrand.

We will show in this paper that it is important to use
a numerical integration scheme that approximates the principal
value integral of a singular integrand when the self energy
convolution is performed directly on the real axis.
We show that failure to do so may lead to significant error
in integration over some frequency regions.
In this study, we address the singularities in the integration that arise from taking the zero
broadening limit in Green's function only.
The approximation to the principal value is obtained by
replacing the numerator of the integrand with a piecewise
polynomial approximation and performing a principal
value integration analytically on the approximate integrand.
We show that taking the zero broadening limit improves the
accuracy of the numerical integration.

When the spectral representations of $G$ and $W$ are used
to derive an expression for $\Sigma$, there are a number of
ways to group different terms. These different groupings lead to different
expressions.  We show that applying the numerical integration
scheme proposed in this paper to different expressions
may lead to different numerical accuracy. For
particular frequency ranges, one expression may be preferred over
the other.

\section{Self energy integration}\label{sec:self-energy-integ}
In the GW approximation, the electron self energy is
expressed as $i$ times the product of the time-ordered one-particle Green's function,
denoted by $G(r,r';t)$, and a screened Coulomb term, denoted
by $W(r,r';t)$, in the time domain. Taking a Fourier
transform with respect to $t$ yields the frequency representation
of the self energy,
\begin{equation}
\Sigma(r,r';\omega) = \frac{i}{2\pi} \int_{-\infty}^{\infty}
d\omega' G(r,r';\omega-\omega')W(r,r',\omega')e^{-i\omega' \eta},
\label{eq:sigma}
\end{equation}
where $G(r,r';\omega)$ and $W(r,r';\omega)$ are obtained from
Fourier transforms of $G(r,r';t)$ and $W(r,r';t)$, respectively.
Note that the $e^{-i\omega' \eta}$ term with a positive infinitesimally small
$\eta$ factor is introduced to ensure the proper convergence of the
Fourier transform. Such a factor should also appear in
the Fourier transforms for $G(r,r';\omega)$ and $W(r,r';\omega)$ themselves.

In practice, this convolution~\eqref{eq:sigma} must be evaluated
numerically.  The accuracy of the numerical integration can have a
significant effect on the quantitative and qualitative behaviors
of the approximate solution to Dyson's equation.

In the frequency domain, Green's function has the spectral
representation
\begin{equation}\label{eq:green}
G(r,r';\omega) = \sum_{j}
\frac{\phi_j(r)\phi^*_j(r')}{\omega-\varepsilon_j+i\eta
\text{sgn}(\varepsilon_j-\mu)}
\end{equation}
where $(\varepsilon_j,\phi_j)$, $j=1,2,...$ are quasi-particle
eigenvalues (energies) and orbitals enumerated in an increasing energy
order.  In a computational procedure, they are chosen to be
approximate solutions to Dyson's equation with $\phi_j$ evaluated at $\omega=\varepsilon_j$
for the self energy $\Sigma$.  The first $n_v$
eigenpairs are called the valence (or occupied)
states, and the remaining ones are referred to as the
conduction (or empty) states. The parameter $\mu$ is the chemical
potential that satisfies the condition
$\varepsilon_{n_v} \leq \mu \leq \varepsilon_{n_v+1}$.

The screened Coulomb interaction $W$ can be expressed as
\begin{equation}
W(r,r';\omega) \equiv \int \epsilon^{-1}(r,r'';\omega) v(r'',r') dr'',
\label{eq:hscr}
\end{equation}
where the dielectric function $\epsilon$, defined within the
random phase approximation \cite{Bohm53,GellMann57}, has the form
\[
\epsilon(r,r';\omega) = \delta(r,r') - \int v(r,r'')\chi_0(r'',r';\omega) dr'',
\]
with $v(r,r') = 1/|r-r'|$ being the bare Coulomb potential,
and $\chi_0(r,r';\omega)$ being the time-ordered Fourier transform of
the independent particle polarizability function (operator) that describes the
linear density response to external potential perturbations. The
analytical expression for the non-interacting $\chi_0(r,r';\omega)$ from
a mean-field solution of the system is
\begin{equation}
\begin{split}
\chi_0(r,r';\omega) = & \frac{1}{2}\sum_{i=1}^{n_v}\sum_{j>n_v}
\phi_{i}(r)\phi_{j}^{\ast}(r) \phi^{\ast}_{i}(r')\phi_{j}(r')\\
& \times \left(\frac{1}{\omega
- \Delta \varepsilon_{i,j}+i\eta}-\frac{1}{\omega + \Delta
\varepsilon_{i,j}-i\eta}\right),
\end{split}
\label{eq:chi}
\end{equation}
where $\Delta\varepsilon_{i,j} = \varepsilon_j - \varepsilon_i \ge 0$.

\subsection{The analytic structure of the integrand}\label{sec:structure}
Because the time-ordered screened Coulomb operator $W$ involves the
inverse of the dielectric operator, the analytic structure of the
integrand in~\eqref{eq:sigma} is not immediately clear. In the
following, we examine the finite dimensional approximation of the
integrand in~\eqref{eq:sigma} in detail,
and discuss how the integral~\eqref{eq:sigma} can be evaluated
numerically.

Let us assume that a proper spatial discretization (e.g., plane-wave
expansion) has been used to represent $G$, $W$ and $\Sigma(\omega)$
as $n \times n$ matrices.  Using the spectral representation of $G$ and $W$,
we can express the integral~\eqref{eq:sigma} \cite{Hedin65,Hybertsen86,Hedin70} as
\begin{equation}\label{eq:sigma-cohsex}
  \begin{split}
    \Sigma(\omega)
    =&- \sum_{j=1}^{n_v} \left(\phi_j \phi_j^*\right)\odot
     W(\omega-\varepsilon_j)\\
     &-\frac{1}{\pi}\sum_{j=1}^n (\phi_j \phi_j^*) \odot
    \int^\infty_{0} d\omega' \frac{(W^{r}(\omega')-W^{a}(\omega'))/(2i)}{\omega-\varepsilon_j-\omega'+i\eta},\\
  \end{split}
\end{equation}
where $W^{r/a} = \left[\epsilon^{r/a}\right]^{-1}V$ are the retarded and
advanced screened Coulomb matrix defined in terms of the retarded and advanced
dielectric matrices $\epsilon^{r/a}$, which are in turn
defined in terms of the retarded and advanced polarizability matrices
\begin{equation}
\begin{split}
\chi_0^{r/a}(\omega)
= &
\frac{1}{2}\sum_{i=1}^{n_v}\sum_{j=n_v+1}^{n} \left(\phi_{i}\odot \overline{\phi_{j}}\right)
\left(\phi_{i}\odot \overline{\phi_{j}}\right)^{\ast} \\
& \times\left(\frac{1}{\omega - \Delta
\varepsilon_{i,j}\pm i\eta}-\frac{1}{\omega + \Delta
\varepsilon_{i,j}\pm i\eta}\right),
\end{split}
\label{eq:chi-ra}
\end{equation}
and the discretized bare Coulomb matrix $V$. We use $\odot$ to denote
element-wise multiplication, $\phi_j^{\ast}$ to denote the conjugate
transpose of a column vector $\phi_j$, and $\overline{\phi_j}$ to denote the conjugate
 of $\phi_j$.  The integral in
\eqref{eq:sigma-cohsex} is performed element-wise. We should also
mention that in a practical calculation, the summation over $j$ may be
truncated so that the total number of unoccupied states, which we will
denote by $n_c$, can be less than $n - n_v$.

The expression given in~\eqref{eq:sigma-cohsex} is a specific formulation of
the self energy. The first term in \eqref{eq:sigma-cohsex} is the so-called
screened exchange (SEX) term and the second term is the Coulomb hole (COH) term.
The integrand in
the second term of~\eqref{eq:sigma-cohsex}
is well behaved in the sense that the integral does not diverge.
Such an expression is
also physically appealing, especially in the static case ($\omega=0$) \cite{Hedin65}.

Without loss of generality, we can write the matrix $\chi_0^r(\omega)$ as
\[
\chi_0^r(\omega) =  M [\Omega(\omega)]^{-1} M^{\ast},
\]
where $M$ is an $n \times n_v n_c$ matrix. Each column of $M$ represents the
element-wise product of a discretized $\phi_{i}$ and $\overline{\phi_{j}}$ pair, for
$i=1,2,...,n_v$, $j = n_v+1,n_v+2,...,n_v+n_c$. The $n_v n_c \times n_v n_c$
diagonal matrix $[\Omega(\omega)]^{-1}$ has elements
\begin{equation}
\frac{1}{2}\left(\frac{1}{\omega - \Delta \varepsilon_{i,j}+i\eta}-\frac{1}{\omega +
\Delta \varepsilon_{i,j}+i\eta}\right)
=\frac{\Delta \varepsilon_{i,j}}{(\omega+i\eta)^2 - \Delta \varepsilon_{i,j}^2}.
\label{eq:diagfrac}
\end{equation}

If we let $V$ be the matrix representation of a discretized unscreened Coulomb
operator, we can write $\epsilon^{r}(\omega)$ as
\begin{equation}\label{eq:eps}
\epsilon^r(\omega) = I - V M [\Omega(\omega)]^{-1} M^{\ast}.
\end{equation}

It follows from the Sherman-Morrison-Woodbury formula \cite{Woodbury50,
Hager89} and some additional algebraic manipulations \cite{Liu13}
that
\begin{equation}\label{eq:invepsr-pole}
[\epsilon^{r}(\omega)]^{-1} = I +
\sum_{\ell=1}^{n_vn_c} \frac{1}{2\tau_\ell} VMs_\ell \left(\frac{1}{\omega - \tau_{\ell}+i\eta} -\frac{1}{\omega + \tau_{\ell}+i\eta}\right) (M s_\ell)^\ast,
\end{equation}
where $\tau_{\ell}=\sqrt{\lambda_\ell}$ with $\lambda_\ell$ being an
eigenvalue of the matrix $D^2+DM^{\ast}VM$, $D$ is a diagonal
matrix with $\Delta \varepsilon_{i,j}$, $i=1,2,...,n_v$, $j=n_v+1,...,n_v+n_c$
on its diagonal, and $s_{\ell}$ is the $\ell$th column of $D^{1/2}U$ with
$U$ being the matrix that contains all eigenvectors of $D^2+D^{1/2}M^{\ast}VM D^{1/2}$.
A similar expression can be obtained for $[\epsilon^{a}(\omega)]^{-1}$.
Consequently,
\begin{equation}\label{eq:iepsr-iepsa}
\begin{split}
&(W^{r}(\omega)-W^{a}(\omega))/(2i) \\
= &
\sum_{\ell=1}^{n_vn_c} \frac{1}{2\tau_\ell} VMs_\ell \left[\frac{\eta}{(\omega + \tau_{\ell})^2+\eta^2} -\frac{\eta}{(\omega - \tau_{\ell})^2+\eta^2}\right] (VM s_\ell)^\ast.
\end{split}
\end{equation}

The expression given by~\eqref{eq:iepsr-iepsa} indicates that all
matrix elements of the integrand of the integration in \eqref{eq:sigma-cohsex}
(in any basis) have the same analytic structure.

In principle, if $\tau_\ell$'s are known, the integration
in \eqref{eq:sigma-cohsex} can be evaluated analytically for all $\omega$.
In the limit of $\eta \rightarrow 0$,
$\eta/[(\omega'\pm \tau_{\ell})^2 + \eta^2]$ becomes a
$\delta$-function centered at $\omega'=\mp \tau_\ell$.  Because
$\tau_{\ell}=\sqrt{\lambda_\ell}$ is nonnegative, the COH term
in~\eqref{eq:sigma-cohsex} simplifies to
\begin{equation}
I_{\text{COH}} = \frac{1}{\pi}\sum_{j=1}^n \left(\phi_j\phi_j^\ast\right) \odot \left[\sum_{\ell=1}^{n_vn_c} \frac{1}{2\tau_\ell} \frac{VMs_\ell (V Ms_\ell)^\ast}{\omega - \varepsilon_j - \tau_\ell}\right] .
\label{eq:exact}
\end{equation}

However, obtaining $\tau_{\ell}$ requires computing all
eigenvalues of an $n_v n_c \times  n_v n_c$ matrix, a task that
requires $\mathcal{O}(n^6)$ operations even though $\Sigma$ and
$\epsilon^{r/a}$ are $n \times n$ matrices \cite{Casida95,Tiago06,Lischner12}.
This can be costly
for large systems that contain many atoms.
An alternative way to evaluate the integral in~\eqref{eq:sigma-cohsex}
numerically is to evaluate the integrand at multiple values of
$\omega'$ and use an appropriate quadrature rule to sum up these
function evaluations. In this approach, the
$\epsilon^{r/a}(\omega')$ matrices must be evaluated at a number of
frequencies. Each evaluation is expensive, because it
requires computing $\chi_0^{r/a}(\omega')$ for each $\omega'$,
multiplying $\chi_0^{r/a}(\omega')$ with $V$ and inverting the product.
The complexity of the evaluation is $\mathcal{O}(n^4)$ for each $\omega'$,
where $n$ scales linearly with the number of atoms in the system. Therefore,
we would like to minimize the number of such evaluations as much as
possible without sacrificing the accuracy of the integration.

\subsection{Separating the low and high frequency regions}\label{sec:seperate-region}
Let
$S_\eta(\omega')=(W^{r}(\omega')-W^{a}(\omega'))/(2i)$.
In frequency regions where the poles of $S_\eta(\omega')$ are near
$\omega-\varepsilon_j$, the direct numerical integration
of~\eqref{eq:sigma-cohsex} may be difficult due to the presence of
singularities in the integrand.  As a result, more quadrature points
need to be placed near the poles of $S_\eta(\omega')$.
However, in the tail regions of $1/(\omega - \varepsilon_j - \omega' + i\eta)$,
that are far away from these poles, the Lorentzians in
$S_\eta(\omega')$ should decay to zero rapidly.
Fewer quadrature points are thus needed to approximate
the COH term in that region by a weighted sum. In particular,
we can use a Gauss quadrature rule, i.e.
\begin{equation}
\int_{a}^b d\omega' \frac{S_\eta(\omega')}{\omega-\varepsilon_j-\omega'+i\eta}  \approx
\sum_{i=1}^P  \frac{S_\eta(\omega'_i)}{\omega-\varepsilon_j-\omega'_i+i\eta} \nu_i,
\label{eq:gaussqt}
\end{equation}
where $\omega'_i$ are quadrature points at which the integrand is
evaluated and $\nu_i$ are the properly chosen weights. Typically,
ten or twenty quadrature points are sufficient to produce a highly
accurate integral.

For a given $\omega$, we can apply~\eqref{eq:gaussqt} to
the high frequency interval $[\xi, \infty)$ where
$\xi = \max(\omega - \varepsilon_1, \max_\ell \tau_{\ell}) + \zeta$,
for some modest constant $\zeta \geq 0$. Here $\tau_{\ell},~\ell=1,2,\cdots,n_v n_c$ are the position of poles of $S_\eta(\omega')$ for $\omega'\ge 0$
as shown in \eqref{eq:iepsr-iepsa}.
If we view the matrix $D M^{\ast} VM$
as a small perturbation to the diagonal matrix $D^2$ whose diagonal elements
are $(\Delta \varepsilon_{i,j})^2$, the poles of
$S_\eta(\omega')$ should not be too far away from
$\pm \Delta \varepsilon_{i,j}$.  Hence we can simply estimate
$\max_\ell \tau_{\ell}$ by $\varepsilon_{\max} - \varepsilon_1$, where
$\max= n_v+n_c$. Because
the $\omega$'s of interest are often within $[\varepsilon_1, \varepsilon_{\max}]$,
choosing $\xi = \varepsilon_{\max} - \varepsilon_1 + \zeta$ as the starting point
of the high frequency region is not unreasonable.
Below this region, i.e., within the interval
$(0,\varepsilon_{\max} - \varepsilon_1)$, a different
numerical integration strategy that accounts for the singular
nature of the integrand must be used. We will describe
that strategy in the next section.

\begin{figure}[htbp]
\centering
\includegraphics[width=0.8\textwidth]{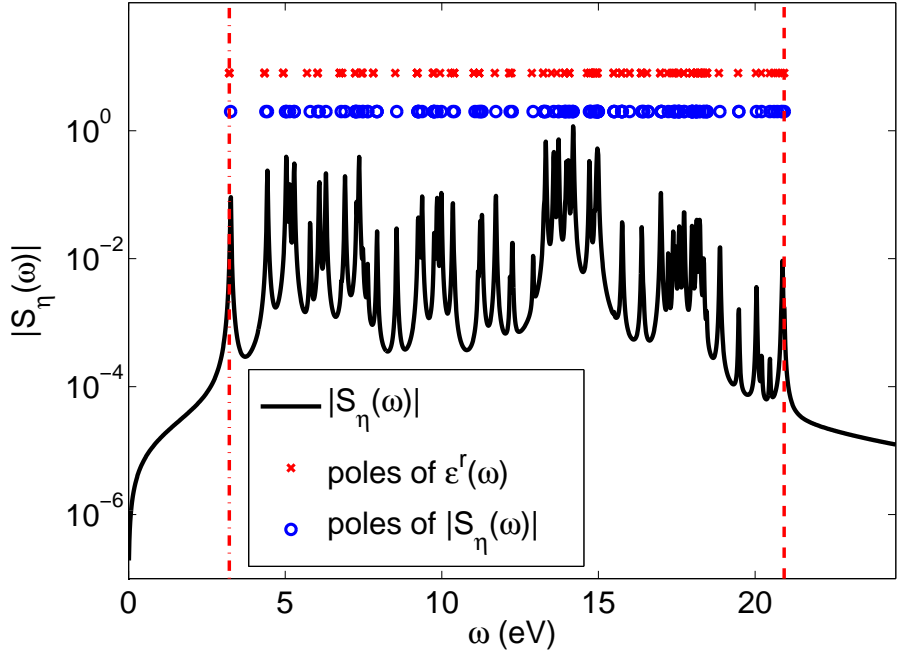}
\caption{\small The diagonal element of $|S_\eta(\omega)|$ of SiH4,
associated with $G=G'=[2~ 0 ~0]$, is shown as the black curve. $G$, $G'$ are in
units of the reciprocal basis $2\pi/a_0$ of the supercell containing the molecule with $a_0=10$ Bohr.
The top of the figure shows the positions of the poles associated
with the corresponding entry in the $\epsilon^{r}(\omega)$ matrix
(the red crosses), as well as the positions of poles of $|S_\eta(\omega)|$
(the blue circles).}
\label{fig:poleall}
\end{figure}

To confirm the above observation, we plot both the poles of a typical
matrix element of the $\epsilon^r(\omega)$ and $S_\eta(\omega)$
associated with a SiH4 molecule in Figure~\ref{fig:poleall}. Our calculation
is performed using KSSOLV \cite{Yang09}, which is a MATLAB toolbox for solving
the Kohn-Sham problem.  The plane-wave expansion is used for discretizing
$\epsilon^r(\omega)$ and $S_\eta(\omega)$.
$(\varepsilon_j,\phi_j)$'s are taken to be the Kohn-Sham DFT eigenpairs.
We will use $\varepsilon_{LUMO}$ and $\varepsilon_{HOMO}$ to denote the
lowest unoccupied (empty) and the highest occupied Kohn-Sham single-particle
eigenvalues below, i.e. $\varepsilon_{LUMO} = \varepsilon_{n_v+1}$ and
$\varepsilon_{HOMO} = \varepsilon_{n_v}$.

Figure~\ref{fig:poleall} shows that poles of
$\epsilon^r(\omega)$, which are at $\Delta \varepsilon_{i,j}$, and those of
$S_\eta(\omega)$, which are at $\tau_\ell$, match up pretty well. In particular,
they are both in the domain $[\min \Delta \varepsilon_{i,j},\max \Delta \varepsilon_{i,j}]
=[\varepsilon_{LUMO}-\varepsilon_{HOMO},\varepsilon_{\max}-\varepsilon_1]=[3.2,20.9]$ eV. Outside of this region, the
magnitude of $S_\eta(\omega)$
decreases rapidly to zero.

\subsection{Quadrature for the low frequency region}\label{sec:quadrature}
The fundamental problem we need to solve in order to evaluate
\eqref{eq:sigma-cohsex} efficiently and accurately is to
properly evaluate an integral of the form
\begin{equation}
I_{\eta}(\gamma) = \int_{0}^\infty \frac{f(\omega')}{\omega'-\gamma + i\eta} d\omega'
\label{eq:prototype}
\end{equation}
for $\gamma > 0$, where $f(\omega')$ contains a linear
combination of a number of Lorentzians centered at $\tau_\ell$'s, which
we do not know in advance.

Taking the $\eta \rightarrow 0$ limit in the denominator of the integrand
reduces the integral to
\begin{equation}
I_{\eta}(\gamma) = -i\pi f(\gamma) + \mbox{PV} \int_{0}^\infty \frac{f(\omega')}{\omega'-\gamma}
d\omega',
\label{eq:pv}
\end{equation}
where PV denotes the principal value.

Hence, we shall now focus on the numerical evaluation of the
principal value integral in~\eqref{eq:pv}.

A simple quadrature rule for integrating $f(\omega')/(\omega'-\gamma)$
numerically on the interval $[a,b]$ is the composite trapezoidal rule.
If we let
\[
\omega'_{i} = a + (i-1) h, \quad i = 1,\ldots, N,
\]
where $h = (b-a)/(N-1)$, the trapezoidal rule gives the following approximation:
\begin{equation}\label{eq:trap}
\int^{b}_{a} \frac{f(\omega')}{\omega'-\gamma} d\omega'
=\sum_{i=1}^{N-1} \frac{1}{2}\left(\frac{1}{\omega'_i-\gamma}f(\omega'_i)
+\frac{1}{\omega'_{i+1}-\gamma}f(\omega'_{i+1})\right) h.
\end{equation}

However, the trapezoidal rule generally does not converge to
the principal value of integral in~\eqref{eq:pv}, even if $h$ is
chosen to be very small.  To see
this, let us assume that that $\gamma \in (\omega'_{I},\omega'_{I+1})$,
for some $1 \le I \le N-1$.  (If $\gamma = \omega_i'$ for some $i$,
we move $\omega_i'$ slightly away from $\gamma$ to avoid floating
point overflow.)
That is, we assume that $\gamma$ can be close to an integration
point $\omega'_{I}$, but never be exactly equal to $\omega'_{I}$.

If $\gamma-\omega'_I=\alpha h$ for some $0<\alpha \ll 1$,
it follows that the term
\begin{equation}
    \abs{\frac{f(\omega'_{I})h}{\omega'_{I}-\gamma}} = \abs{\frac{f(\omega'_I)}{\alpha}}
    \label{}
\end{equation}
in the summation will dominate over other terms in Eq.~\eqref{eq:trap}
if $\alpha \ll \abs{f(\omega'_{I})}$.  As $\alpha\to 0$, the right hand side
of~\eqref{eq:trap} behaves like $1/\alpha$, and it rapidly approaches
$\infty$ independent of $h$.  No other term with an opposite sign
can offset this large spike.  This undesirable behavior can lead to
sharp artificial peaks in the self energy at some $\omega$ values, as
we will show in the next section.
This issue cannot be addressed by simply using a higher order
numerical integration scheme (such as Simpson's rule) that does
not preserve the principal value of the integral.

One way to mitigate the singularity issue is to resort to an alternative
numerical integration scheme that replaces $f(\omega')$, instead of
the entire integrand, with an approximation, and perform a principal
value integration of the approximated integrand analytically on each interval
$[\omega'_i,\omega'_{i+1}]$.
The simplest approximation of $f(\omega')$ is the piecewise constant
approximation
$f(\omega') \approx [f(\omega'_i) + f(\omega'_{i+1})]/2$ for
$\omega' \in [\omega'_i,\omega'_{i+1}]$.  Such an
approximation leads to the following quadrature rule:
\begin{equation}
\text{PV} \int^{b}_{a} \frac{f(\omega')}{\omega'-\gamma} d\omega'
\approx \sum_i \frac{1}{2}\left[f(\omega'_i)+f(\omega'_{i+1})\right] \text{log}
\left|\frac{\omega'_{i+1}-\gamma}{\omega'_{i}-\gamma}\right|.
\label{eq:splitmp}
\end{equation}

A more accurate approximation is a piecewise linear approximation of
$f(\omega')$, which leads to the following quadrature rule:
\begin{equation}
\begin{split}
\text{PV} \int^{b}_{a} \frac{f(\omega')}{\omega'-\gamma} d\omega'
\approx &
\sum_i \left[f(\omega'_{i+1})-f(\omega'_{i})\right.\\
& \left.+\left(f(\omega'_i)+\frac{f(\omega'_{i+1})-f(\omega'_i)}{\omega'_{i+1}-\omega'_i}(\gamma-\omega'_i)\right)
\text{log} \left|\frac{\omega'_{i+1}-\gamma}{\omega'_{i}-\gamma}\right|\right].
\end{split}
\label{eq:splitlp}
\end{equation}

If we denote the absolute error made in~\eqref{eq:splitmp}
and~\eqref{eq:splitlp} by $E_c$ and $E_l$ respectively, it is not difficult to
show \cite{Liu13} that
\begin{eqnarray}
E_c & \le & C_c h ||f'||_{\infty} \left(1 + \abs{\log h}+ \abs{\log \alpha} +
    \abs{\log (1-\alpha)}\right), \label{eq:thm-mp}\\
E_l & \le & C_l h^2 ||f''||_{\infty} (1 + \abs{\log h}), \label{eq:thm-lp}
\end{eqnarray}
for some constants $C_c$ and $C_l$ that are independent of $h$.
These error bounds indicate that the accuracy of the quadrature should
improve as we decrease $h$.

In~\eqref{eq:splitmp} consider $\gamma-\omega'_I=\alpha h$ with $0<\alpha \ll 1$ and the term
\begin{equation}\label{eq:Iterm}
    \begin{split}
    &\frac{1}{2}\left( f(\omega'_{I-1})+f(\omega'_{I}) \right) \log \abs{\omega'_{I}-\gamma} -
    \frac{1}{2}\left( f(\omega'_{I})+f(\omega'_{I+1}) \right) \log \abs{\omega'_{I}-\gamma}\\
    &= \frac12 \left( f(\omega'_{I-1})-f(\omega'_{I+1}) \right) \log (\alpha h).
    \end{split}
\end{equation}
This term becomes the dominating term if
$\alpha \ll \exp\{-\frac{1}{\abs{f(\omega'_{I-1})-f(\omega'_{I+1})}}\}/h$.

Note that $\abs{f(\omega'_{I-1})-f(\omega'_{I+1})}\sim O(h)$, and that
$\lim_{h\to 0} e^{-C/h}/h = 0$. Thus, for a given $\alpha$, we can always
find an $h$ small enough such that~\eqref{eq:Iterm} does not become
the dominating term.  As a result, even if we replace $f(\omega')$
by a piecewise constant, the corresponding quadrature is more stable
than the standard trapezoid rule. Using the piecewise linear approximation
further improves the accuracy of the numerical integration without
incurring additional function evaluation cost.

An alternative way to overcome the difficulty with the singularity
is to keep the parameter $\eta$ finite in the denominator of the integrand
in~\eqref{eq:prototype}. In this case, we can also replace $f(\omega')$
with a piecewise polynomial (or spline)
approximation and integrate the approximate integrand analytically
on each interval. For example, if we approximate $f(\omega')$ by
a piecewise linear function, the quadrature rule, which is used in
\cite{Shishkin06}, becomes
\begin{equation}\label{eq:kresse}
  \begin{split}
 &\int^{b}_{a} \frac{f(\omega')}{\omega'-\gamma+i\eta} d\omega'  \\
 = & \sum_i \left[\frac{f(\omega'_i)}{\omega'_i-\omega'_{i-1}} \int^{\omega'_{i}}_{\omega'_{i-1}}
     \frac{\omega'-\omega'_{i-1}}{\omega'-\gamma+i\eta}d\omega' + \frac{f(\omega'_{i+1})}{\omega'_{i}-\omega'_{i+1}} \int^{\omega'_{i+1}}_{\omega'_i}
     \frac{\omega'-\omega'_{i+1}}{\omega'-\gamma+i\eta}d\omega'\right].
  \end{split}
\end{equation}
The integrals within the square brackets above can be evaluated analytically. To simplify
our discussion, we do not give the analytical expression here, which is slightly more complicated than
the $\log$ function that appears in~\eqref{eq:splitlp}.

Our numerical examples in the next section show that
the quadrature rule based on~\eqref{eq:kresse} is not as accurate as
the one based on \eqref{eq:pv} and \eqref{eq:splitlp}. The effect of the broadening parameter
is visible near at least some frequencies.

\subsection{Different explicit forms: full-frequency COHSEX vs. XCOR}\label{sec:region}
We should mention that applying the numerical integration scheme to the full-frequency COH term in
\eqref{eq:sigma-cohsex} and evaluating the SEX term directly by
computing $W(\omega - \varepsilon_j)$ according to~\eqref{eq:hscr}
can create a potential numerical issue. To see this, we express
$W(\omega-\varepsilon_j)$ in terms of its spectral representation
\begin{equation}
W(\omega-\varepsilon_j) = V - \frac{1}{\pi} \int_{0}^{\infty} d\omega'
S_\eta(\omega') \left[
\frac{1}{\omega-\varepsilon_j-\omega'+i\eta} - \frac{1}{\omega-\varepsilon_j+\omega'-i\eta}\right].
\label{eq:spectralw}
\end{equation}

Under this representation, the SEX term can be seen to contain the component
\begin{equation}
\label{eq:occterm}
\frac{1}{\pi}\sum_{j=1}^{n_v} (\phi_j\phi_j^{\ast})
\odot \int_{0}^{\infty} d\omega' \frac{S_\eta(\omega')}
{\omega-\varepsilon_j-\omega'+i\eta},
\end{equation}
which cancels with the summation over the occupied states in the COH term.
After the cancellation, what is left can be written as
\begin{equation}
\label{eq:sigma-xcor}
\begin{split}
\Sigma(\omega) = & -\sum_{j=1}^{n_v} (\phi_j\phi_j^{\ast}) \odot V
  -\frac{1}{\pi}\sum_{j=1}^{n_v} (\phi_j\phi_j^{\ast}) \odot \int_{0}^{\infty} d\omega'
  \frac{S_\eta(\omega')}
  {\omega-\varepsilon_j+\omega'-i\eta} \\
  &-\frac{1}{\pi}\sum_{j=n_v+1}^{n} (\phi_j\phi_j^{\ast}) \odot \int_{0}^{\infty} d\omega' \frac{S_\eta(\omega')}
  {\omega-\varepsilon_j-\omega'+i\eta}.
\end{split}
\end{equation}
We call this expression for the self energy the XCOR expression because the first term
of the expression is exactly the exchange (X) term, and what is left over can be viewed
as the correlation (COR) term.

The XCOR and COHSEX expressions are mathematically equivalent.
They correspond to different ways of grouping different terms
when spectral representations of $G$ and $W$ are substituted into
\eqref{eq:sigma}. However, because the denominator of the first
integrand in \eqref{eq:sigma-xcor} is slightly different from that
in \eqref{eq:sigma-cohsex}, and the single-particle states over
which the summations are performed are also different in these
expressions, numerical integration can give different results.
Depending on the $\omega$ value of interest,
it may be more advantageous to numerically integrate one expression
than the other.
\begin{figure}[ht]
\centering
\includegraphics[width=0.65\textwidth]{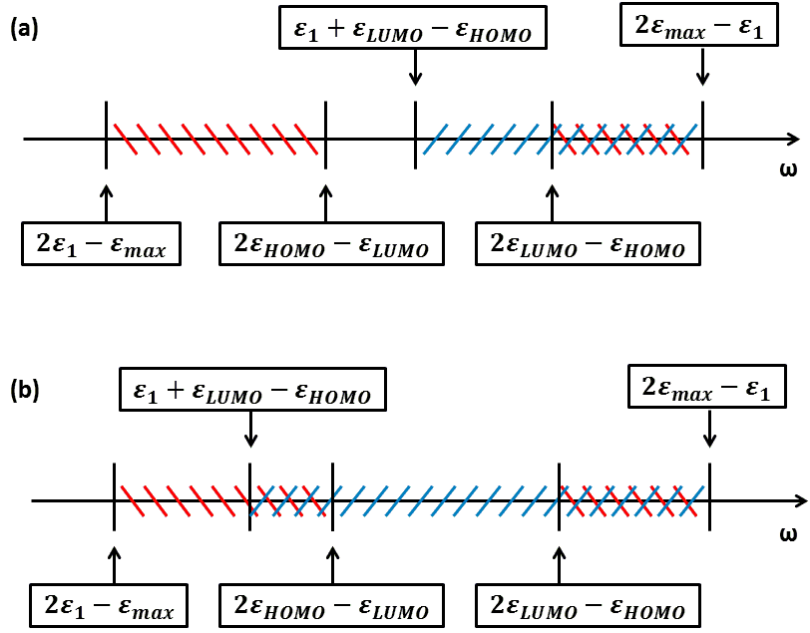}
\caption{The regions of $\omega$ in which the SFC is violated for
the full-frequency COHSEX (blue slashes) and XCOR (red backslashes) expressions.
In (a), we assume $2\varepsilon_{HOMO}-\varepsilon_{LUMO} <
\varepsilon_1+\varepsilon_{LUMO}-\varepsilon_{HOMO}$. In (b), we
assume $2\varepsilon_{HOMO}-\varepsilon_{LUMO} >
\varepsilon_1+\varepsilon_{LUMO}-\varepsilon_{HOMO}$.}
\label{fig:region}
\end{figure}
More precisely, we should choose an expression in which the centers of
the Lorentzians in the numerator ($S_\eta(\omega')$) of the integrand do not
lie in a region that contains $\omega'$ values that make
the denominator nearly zero. We will refer to this criterion
as the {\em singularity-free criterion} (SFC).

For simplicity, from now on, we use the following
notations for different groups of indexes. Let $I_{occ}=\{1,2,\cdots,n_v\}$, $I_{emp}=\{n_v+1, n_v+2,\cdots, \max\}$,
$I_{all}=\{1,2,\cdots,\max\}$, and $I_{\ell}=\{1,2,\cdots, n_v n_c\}$.
Because the centers of the Lorentzians
in the numerators, denoted by $\tau_{\ell}$'s
($\ell\in I_{\ell}$), are the positions of the poles of $\left[\epsilon^{r/a}\right]^{-1}$,
which roughly lie in the region
$[\varepsilon_{LUMO}-\varepsilon_{HOMO},\varepsilon_{\max}-\varepsilon_1]$
as we explained in Section \ref{sec:seperate-region},
we can estimate regions of $\omega$ in which the SFC is violated for
both the COHSEX and the XCOR expressions.
The integrand in the COHSEX expression \eqref{eq:sigma-cohsex} has poles near
$\omega-\varepsilon_j (j\in I_{all})$.
Thus, the SFC is violated when $\omega \approx \varepsilon_j + \tau_{\ell}$.
Hence, the region in which SFC is violated can be estimated by
\begin{equation}\label{eq:reg-cohsex}
\left[\min_{j\in I_{all} \atop  \ell\in I_{\ell}}
\{\varepsilon_j + \tau_{\ell}\}, \max_{j\in I_{all} \atop \ell\in I_{\ell}}
\{\varepsilon_j + \tau_{\ell}\}\right]
\approx \left[\varepsilon_1+\varepsilon_{LUMO}-\varepsilon_{HOMO},2\varepsilon_{\max}-\varepsilon_1\right].
\end{equation}
This region
is marked by blue slashes in Figure~\ref{fig:region}.
A similar analysis shows that the integrand in \eqref{eq:sigma-xcor}
has singularities near
$\omega \approx \varepsilon_j - \tau_{\ell} (j\in I_{occ})$ due to
the second term in \eqref{eq:sigma-xcor} and near
$\omega \approx \varepsilon_j + \tau_{\ell}(j\in I_{emp})$
due to the third term. Hence, the regions in which the SFC is violated
for the XCOR expression can be estimated by
\begin{equation}\label{eq:reg-xcor}
\begin{split}
&\left[\min_{j\in I_{occ} \atop  \ell\in I_{\ell}}
\{\varepsilon_j - \tau_{\ell}\}, \max_{j\in I_{occ} \atop  \ell\in I_{\ell}}
\{\varepsilon_j - \tau_{\ell}\}\right]\bigcup \left[\min_{j\in I_{emp} \atop  \ell\in I_{\ell}}
\{\varepsilon_j + \tau_{\ell}\}, \max_{j\in I_{emp} \atop  \ell\in I_{\ell}}
\{\varepsilon_j + \tau_{\ell}\}\right]\\
= & \left[\varepsilon_1-\max \tau_{\ell}, \varepsilon_{HOMO}
-\min \tau_{\ell}\right]\bigcup \left[\varepsilon_{LUMO}+\min \tau_{\ell}, \varepsilon_{\max}+\max \tau_{\ell}\right]\\
\approx & \left[2\varepsilon_1-\varepsilon_{\max}, 2\varepsilon_{HOMO}
-\varepsilon_{LUMO}\right]\bigcup \left[2\varepsilon_{LUMO}-\varepsilon_{HOMO}, 2\varepsilon_{\max}-\varepsilon_1\right].
\end{split}
\end{equation}
These regions are marked by the red backslash symbols in Figure~\ref{fig:region}.

Since $\varepsilon_1+\varepsilon_{LUMO}-\varepsilon_{HOMO}$ is not always
greater (or less) than $2\varepsilon_{HOMO}-\varepsilon_{LUMO}$,
two scenarios must be considered. They are shown in Figure~\ref{fig:region} (a) and (b) seperately.
Clearly, Figure~\ref{fig:region} shows that the XCOR expression is always better than the COHSEX expression
in the frequency region
$[2\varepsilon_{HOMO}-\varepsilon_{LUMO}$, $2\varepsilon_{LUMO}-\varepsilon_{HOMO}]$,
because in this region the SFC always holds for the XCOR expression but
not necessarily for COHSEX.  Thus, even the trapezoidal rule may
work reasonably well for the XCOR expression. The COHSEX expression is
always preferred in $\left[2\varepsilon_1-\varepsilon_{\max},\varepsilon_1+\varepsilon_{LUMO}-\varepsilon_{HOMO}\right]
$.
We can also see that in
$
\left[2\varepsilon_{LUMO}-\varepsilon_{HOMO},2\varepsilon_{\max}-\varepsilon_1\right],
$
the SFC is violated for both the COHSEX and XCOR expressions. Therefore,
in this region (where the blue slashes intersect with the red blackslashes
in Figure~\ref{fig:region}), the use of the trapezoidal rule is likely
to result in large errors regardless whether the COHSEX or XCOR expression
is used.

\section{Numerical example}\label{sec:numer}
In this section, we demonstrate the advantage of using the
principal value integration based on quadrature rule such as~\eqref{eq:splitlp}
to perform the self energy integration.
We will first illustrate this point by using a simple model test problem
in section~\ref{sec:model} of which we know the exact solution.
We then show the effects of using different numerical integration schemes
on the full-frequency COHSEX and XCOR expressions of the self energy for a small molecule.
These tests are performed by modifying and running
the BerkeleyGW software package \cite{Deslippe12}.

\subsection{Model test problem} \label{sec:model}
In the simple model test problem, we choose $f(\omega')$
in~\eqref{eq:prototype} to be a single Lorentzian centered at 0, i.e., we let
\[
f(\omega') = \frac{\eta}{\omega'^2 + \eta^2},
\]
with $\eta$ set to 0.1, and evaluate
\begin{equation}\label{eq:model}
I(\gamma) = \int^{10}_{-10} \frac{f(\omega')}{\omega'-\gamma}d\omega',
\end{equation}
for a set of $\gamma$ values within $[-1,1]$.
The analytical solution to this integration problem is
\[
\begin{split}
I(\gamma) =& -\frac{1}{2(\gamma^2+\eta^2)}\left[\eta \log\left|\frac{(10+\gamma)^2+\eta^2}{(-10+\gamma)^2+\eta^2}\right|\right.\\
           & \left. + 2\gamma\left(\arctan\left(\frac{10+\gamma}{\eta}\right)-
           \arctan\left(\frac{-10+\gamma}{\eta}\right)\right)\right].
\end{split}
\]
Hence, we can compute the errors associated with different numerical
integration schemes.

In all our numerical integration schemes, we choose the integration step
size to be $h = \eta = 0.1$.

\begin{figure}[ht]
\centering
\includegraphics[width=0.48\textwidth]{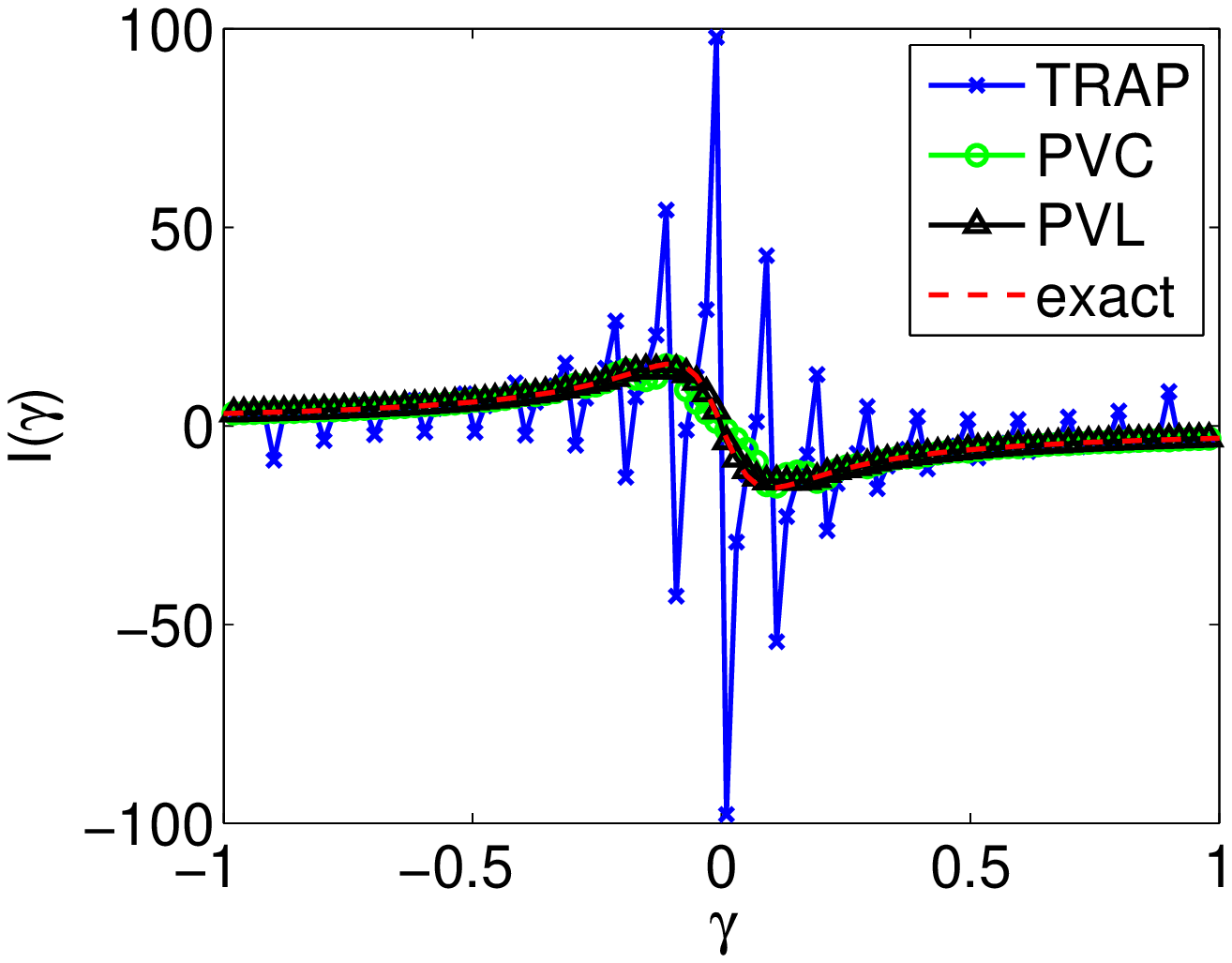}
\includegraphics[width=0.48\textwidth]{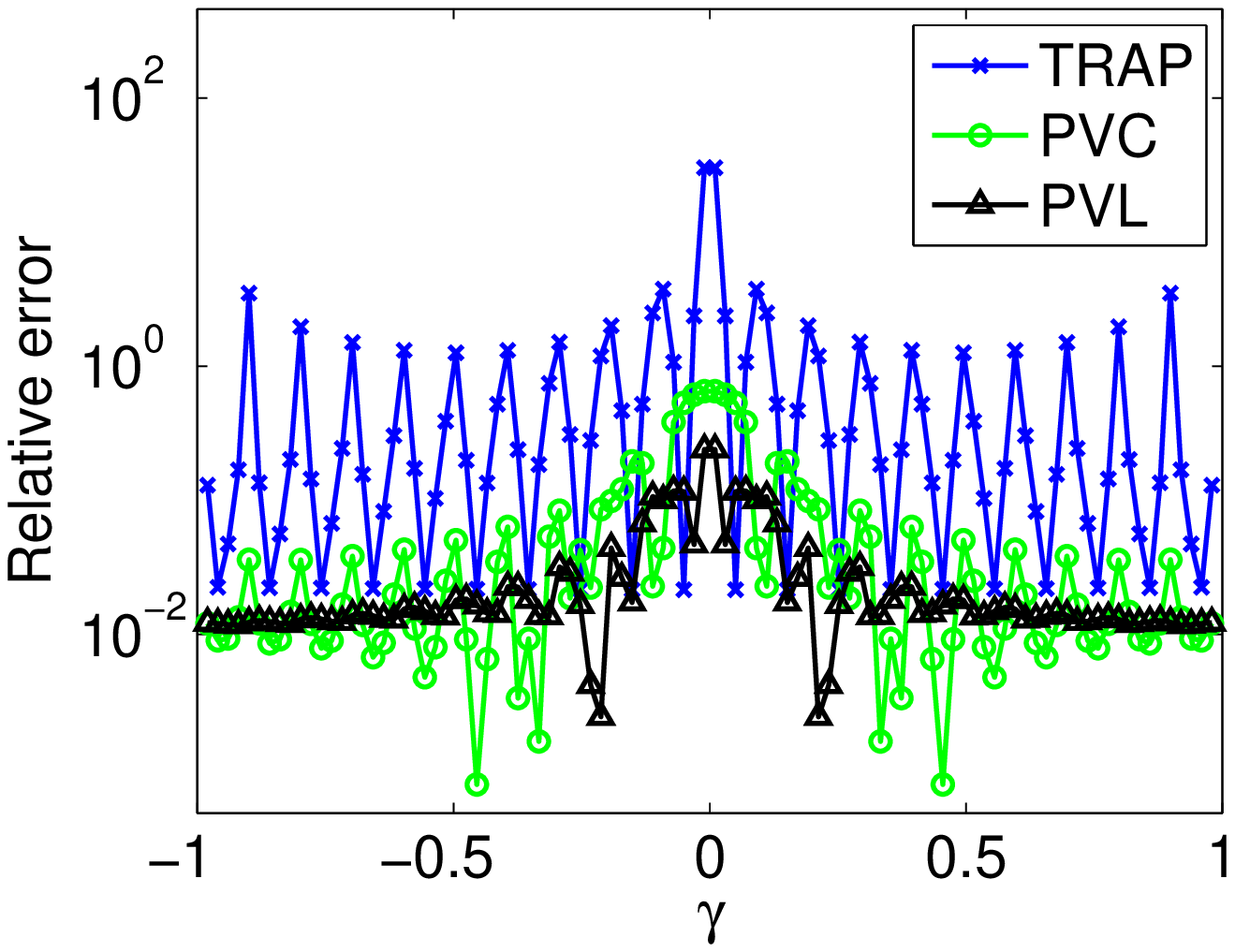}
\caption{The results produced by the trapezoidal rule (TRAP) applied
directly to~\eqref{eq:model} and alternative quadrature rules that perform
a principal value integration analytically after $f(\omega')$ is
approximated by a piecewise constant function (PVC) and a piecewise linear function (PVL).
Left: the analytic (``exact") integral $I(\gamma)$ and the
numerical approximations of $I(\gamma)$. Right: the relative errors.}\label{fig:mod0} %\vskip 0.2cm
\end{figure}

In Figure \ref{fig:mod0}, the exact $I(\gamma)$, the approximation
obtained from different numerical integration schemes
(\eqref{eq:trap}, \eqref{eq:splitmp} and \eqref{eq:splitlp})
and their corresponding relative errors are shown.
We can clearly see that the relative errors associated with the
trapezoidal rule can be two orders of magnitude larger than those
associated with the alternative quadrature rules~\eqref{eq:splitmp} and
\eqref{eq:splitlp}.

We observe that the largest error occurs near (but not at) $\gamma=0$ where
the Lorentzian has a relatively large value, and the nearest two integration
points are not symmetric with respect to $\gamma$.  This observation
is consistent with the analysis in section~\ref{sec:quadrature}.

\subsection{The self energy of methane}\label{sec:BGW}
We now show how different integration schemes perform for
the self energy calculation of a methane molecule.
We implemented \eqref{eq:trap}, \eqref{eq:splitmp},
\eqref{eq:splitlp}, and \eqref{eq:kresse} in the BerkeleyGW software \cite{Deslippe12}.
We applied these integration schemes to both the full-frequency COHSEX and
XCOR self energy expressions. The abbreviations for
these schemes and the corresponding legends we use for plotting
are listed in Table~\ref{tab:quadlabels}.
\begin{table}[ht]
\begin{small}
\begin{center}
%\leavevmode
\begin{tabular}{|p{2.5cm}|p{6cm}|p{3cm}|}
  \hline
  Label      &  Integration scheme & Legend for plotting\\ \hline \hline
 COHSEX-TRAP &  Trapezoidal rule applied to the full-frequency COHSEX expression & cyan line (square) \\ \hline
 COHSEX-PVC  &  Replace the numerator of the COH term by a piecewise
constant approximation and perform principal value integration analytically~\eqref{eq:splitmp}& black dash line
 (cross)\\ \hline
 COHSEX-PVL  &  Replace the numerator of the COH term by a piecewise linear
approximation and perform principal value integration analytically~\eqref{eq:splitlp}  & blue line (circle)\\ \hline
% COHSEX-PVL $(\eta=0.1)$ &  The quadrature rule \eqref{eq:kresse} applied to the COHSEX expression with $\eta=0.1$& green dash dot line \\ \hline
 XCOR-TRAP   &  Trapezoidal rule applied to the XCOR expression & green line (diamond) \\ \hline
 XCOR-PVC    &  Replace the numerator of the COR term by a piecewise
constant approximation and perform principal value integration analytically~\eqref{eq:splitmp} & blue dash line (circle)\\ \hline
 XCOR-PVL    &  Replace the numerator of the COR term by a piecewise linear
approximation and perform principal value integration analytically~\eqref{eq:splitlp} & black line (cross)\\\hline
% XCOR-PVL $(\eta=0.1)$ &  \eqref{eq:kresse} applied to the XCOR expression with $\eta=0.1$ & cyan dash dot line \\ \hline
\end{tabular}
\caption{Labels for different numerical integration schemes tested.} \label{tab:quadlabels}
\end{center}
\end{small}
\end{table}

We compare our results with an analytical integration scheme
based on~\eqref{eq:exact} \cite{Casida95,Tiago06,Lischner12}. Because the analytical integration
scheme takes the $\eta \rightarrow 0$ limit in both the numerator
and denominator of the integrand, and introduces broadening later on for
plotting, our numerical integration results will
never be exactly the same as analytical results. However, a numerically
accurate integration scheme should be close to the analytical result, and
should not exhibit unexpected features (e.g., peaks) not present in the
analytical results.

We use the plane-wave
density-functional theory (DFT) program Quantum Espresso \cite{Giannozzi09}
to compute the ground-state Kohn-Sham eigenvalues $\varepsilon_j$ and
wavefunctions $\phi_j$ of a methane molecule placed in a supercell of
size $16\times 16\times 16~\text{Bohr}^3$.  The Perdew-Burke-Ernzherhof
(PBE) approximation \cite{Perdew96}
to the exchange-correlation functional, and Troullier-Martins
norm-conserving pseudopotentials \cite{TroullierM91}
are used in the DFT calculation. The plane-wave kinetic energy cutoff used
in the calculation is 90 Ry.
The self energy is approximated by the so-called $G_0W_0$ scheme in which
both Green's function \eqref{eq:green} and the screened Coulomb term
\eqref{eq:hscr} are constructed from the ground-state Kohn-Sham eigenpairs.
The polarizability, dielectric function
and self energy are all calculated using the BerkeleyGW software package
\cite{Deslippe12}.
In the polarizability calculation, we truncate the empty states summation
and use only $n_c = 52$ conduction (empty) states. The number of valence
(occupied) states is $n_v=4$. We employ a 5.0 Ry dielectric matrix
truncation cutoff and truncate the bare Coulomb in the real space
beyond the 10 Bohr radius. It should be pointed out that
this is not a converged calculation considering the small
number of empty states and $(G,G')$ pairs determined by the dielectric
matrix truncation.
The purpose of this calculation is merely to
analyze and compare different numerical integration schemes.

The broadening parameter $\eta$ is set to 0.1 eV.
As we explained in the previous section, the numerical integration
of~\eqref{eq:sigma} is performed separately on two regions.
In the low frequency region $[0.0,30.0]$ eV, the integration is
performed by using a uniform grid with 0.1 eV spacing between the grid points.
This is the region where all of the poles of the
$S_\eta(\omega')$ lie.
A coarser grid is chosen for the integration performed on $[30.0,120.0]$ eV
because the integrand varies slowly with respect to $\omega'$ in this
region and its magnitude is relatively small.  Beyond $120.0$ eV, the
integrand is negligibly small so that the integral of the tail can be ignored.

\begin{figure}[htbp]
\centering
\includegraphics[width=0.8\textwidth]{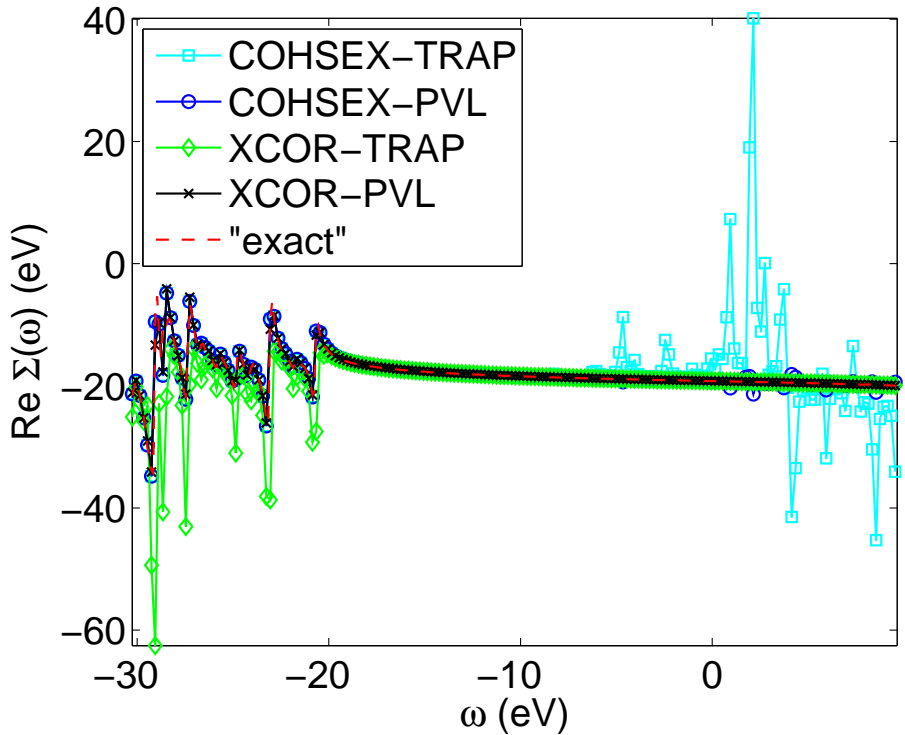}
\caption{The real part of $\langle\phi_4|\Sigma(\omega)|\phi_4\rangle$ for
a methane molecule computed by four different numerical integration
schemes.}
\label{fig:ch4_sigma}
\end{figure}

In Figure \ref{fig:ch4_sigma}, we plot the real part of the HOMO
component of $\Sigma(\omega): \langle\phi_4|\Sigma(\omega)|\phi_4\rangle$ where
the vacuum correction $-0.254$ eV is included, for a
number of $\omega$ values between $-30$ eV and $10$ eV. Four different
integration schemes are used to generate the plot. As a reference,
we also show the self energy values computed from the analytic
expression~\eqref{eq:exact},
which we consider to be ``exact".  This figure shows that the use of principal
value integration is the key to maintaining the accuracy of the numerical
integration scheme regardless whether the COHSEX or XCOR expression
of the self energy is used.  If we simply apply the trapezoidal
rule to integrate the COHSEX expression, large errors are observed
in the frequency region $[$-8, 10$]$ eV.  Similarly, large error is observed
in the $[$-30, -20$]$ eV region when the trapezoidal rule is
applied to the XCOR expression.  This observation is consistent
with the analysis given in section~\ref{sec:region}. Using the
parameters given in Table~\ref{tab:ranges}, we can estimate
the region of $\omega$ in which the SFC is violated for the
COHSEX expression to be
$\left[\varepsilon_1+\varepsilon_{LUMO}-\varepsilon_{HOMO}, 2\varepsilon_{\max}-\varepsilon_1\right] = [$-8.1, 37.0$]$ eV. This interval
contains the interval $[$-8, 10$]$ eV in which large errors are
observed for integrating the COHSEX expression with the trapezoidal rule.
Similarly, the estimated region in which the SFC is violated for
the XCOR expression is $[-43.7,-17.9]\cup[8.2,37.0]$ eV. This is
also consistent with our observation.
\begin{table}[htbp]
\begin{small}
\begin{center}
%\leavevmode
\begin{tabular}{|c|c|c|c|c|c|}
  \hline
  $\varepsilon_1$ & $\varepsilon_{HOMO} $ & $\varepsilon_{LUMO}$ & $\varepsilon_{\max}$ & $\varepsilon_{LUMO}-\varepsilon_{HOMO}$ & $\varepsilon_{\max}-\varepsilon_1$ \\ \hline
   -16.8          & -9.2                  & -0.5                 & 10.1
 & $8.7$ & $26.9$ \\ \hline
\end{tabular}
\caption{The ground-state Kohn-Sham eigenvalues $\varepsilon_j$
(in unit eV) for methane obtained from Quantum Espresso \cite{Giannozzi09}.} \label{tab:ranges}
\end{center}
\end{small}
\end{table}

\begin{figure}[htbp]
\centering
\includegraphics[width=0.45\textwidth]{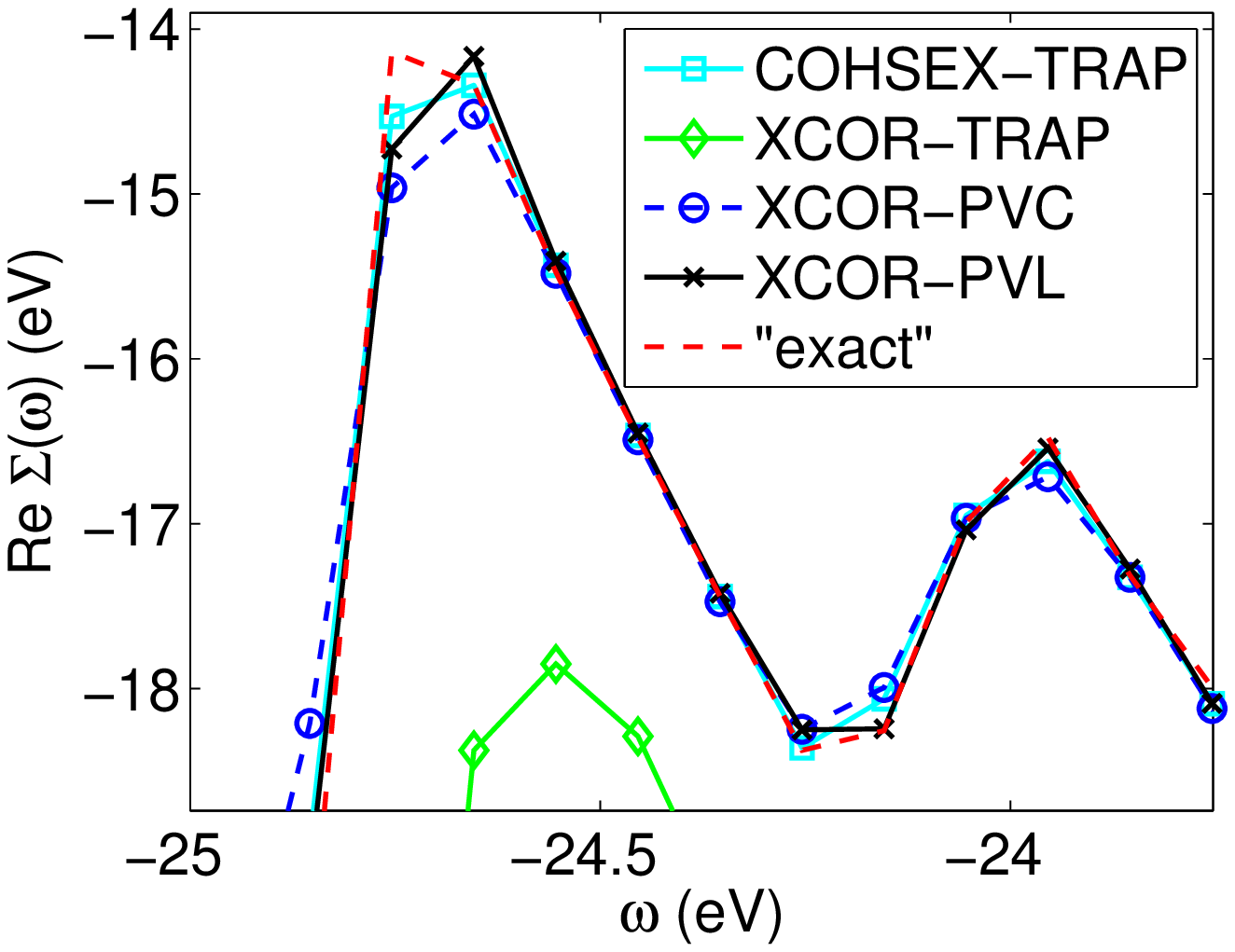}
\includegraphics[width=0.45\textwidth]{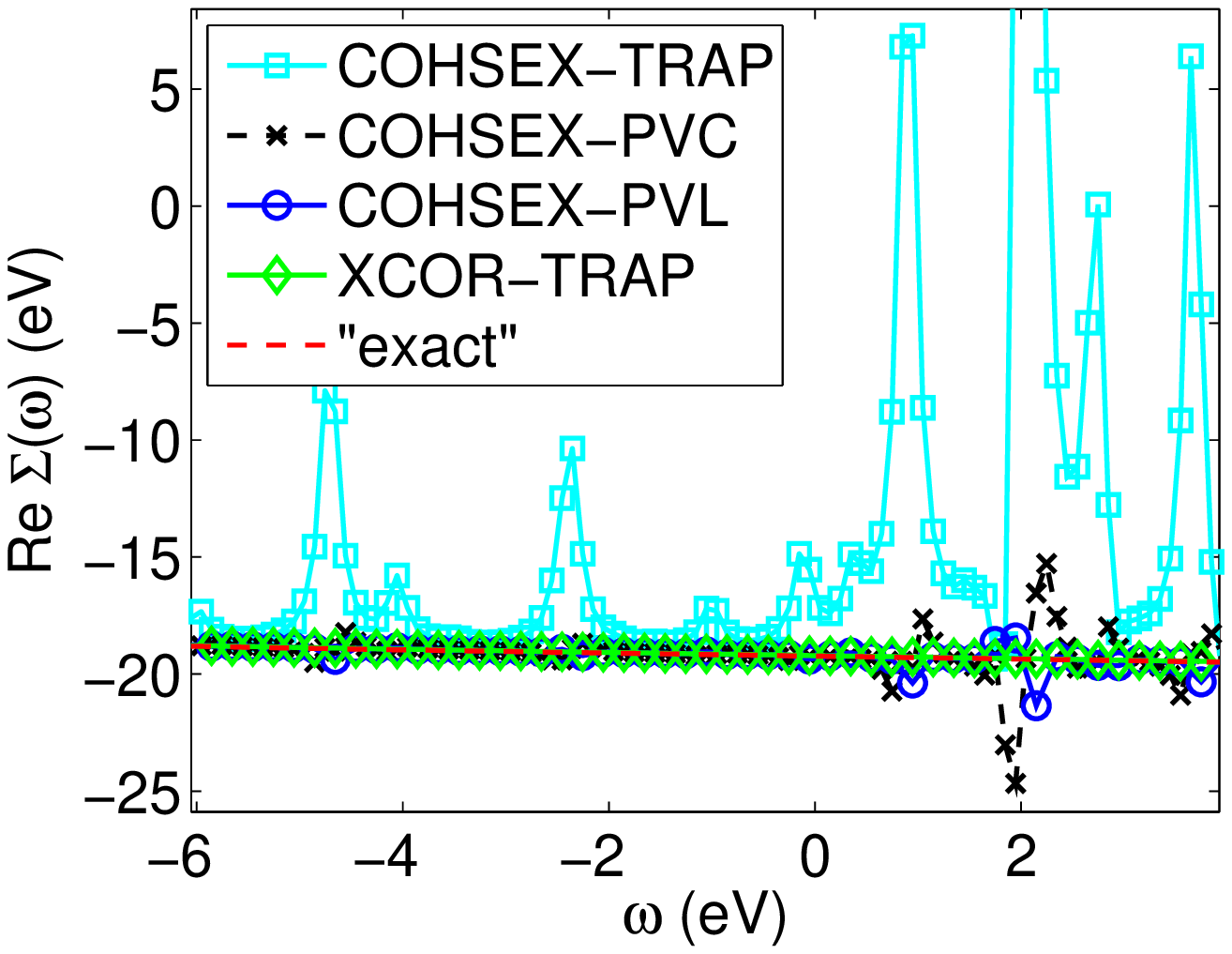}
\caption{The computed real part of $\langle \phi_4|\Sigma(\omega)|\phi_4\rangle$
for a methane molecule where $\omega \in [-25,-23.5]\cup[-6,4]$ eV.}
\label{fig:ch4_sigma_zoom}
\end{figure}

In Figure~\ref{fig:ch4_sigma_zoom}, we zoom in Figure \ref{fig:ch4_sigma} and observe that
the use of principal value integration indeed dramatically reduces
the amount of numerical integration error for both the COHSEX and XCOR expression.
For example, for $\omega \in [-25,-23.5]$ eV, COHSEX is the preferred expression
to integrate because the COH terms does not have any pole in
the region of integration. For $\omega \in [-6, 4]$ eV, XCOR is
clearly the preferred expression to integrate.
However, even if the preferred expression is not used for the
integration, the accuracy of the principal value integration improves
when a better approximation of the numerator (e.g., piecewise linear)
is used. The use of higher order approximation of the numerator also
allows us to use a larger $h$ and fewer function evaluations.

\begin{figure}[htbp]
\centering
\includegraphics[width=0.45\textwidth]{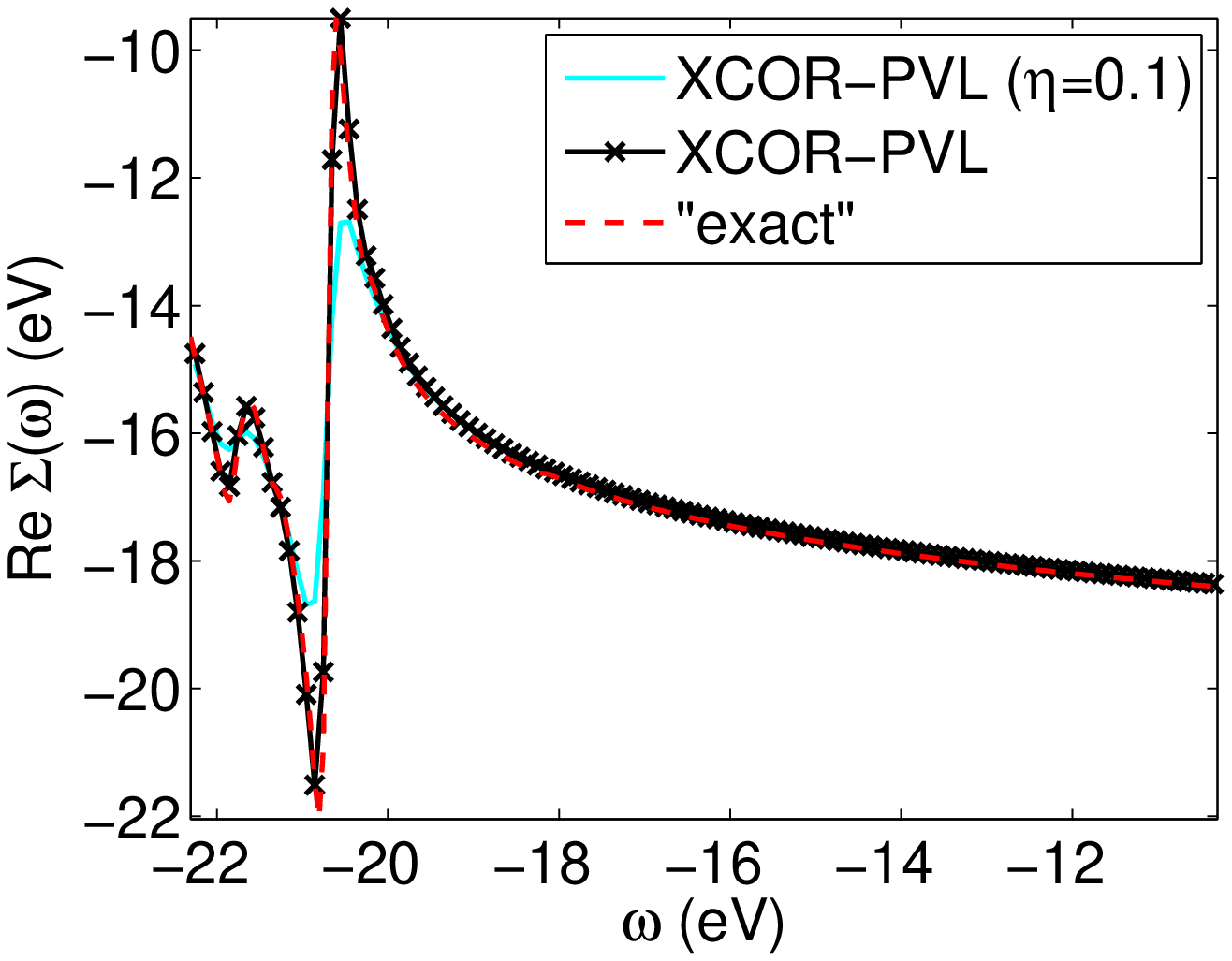}
\includegraphics[width=0.45\textwidth]{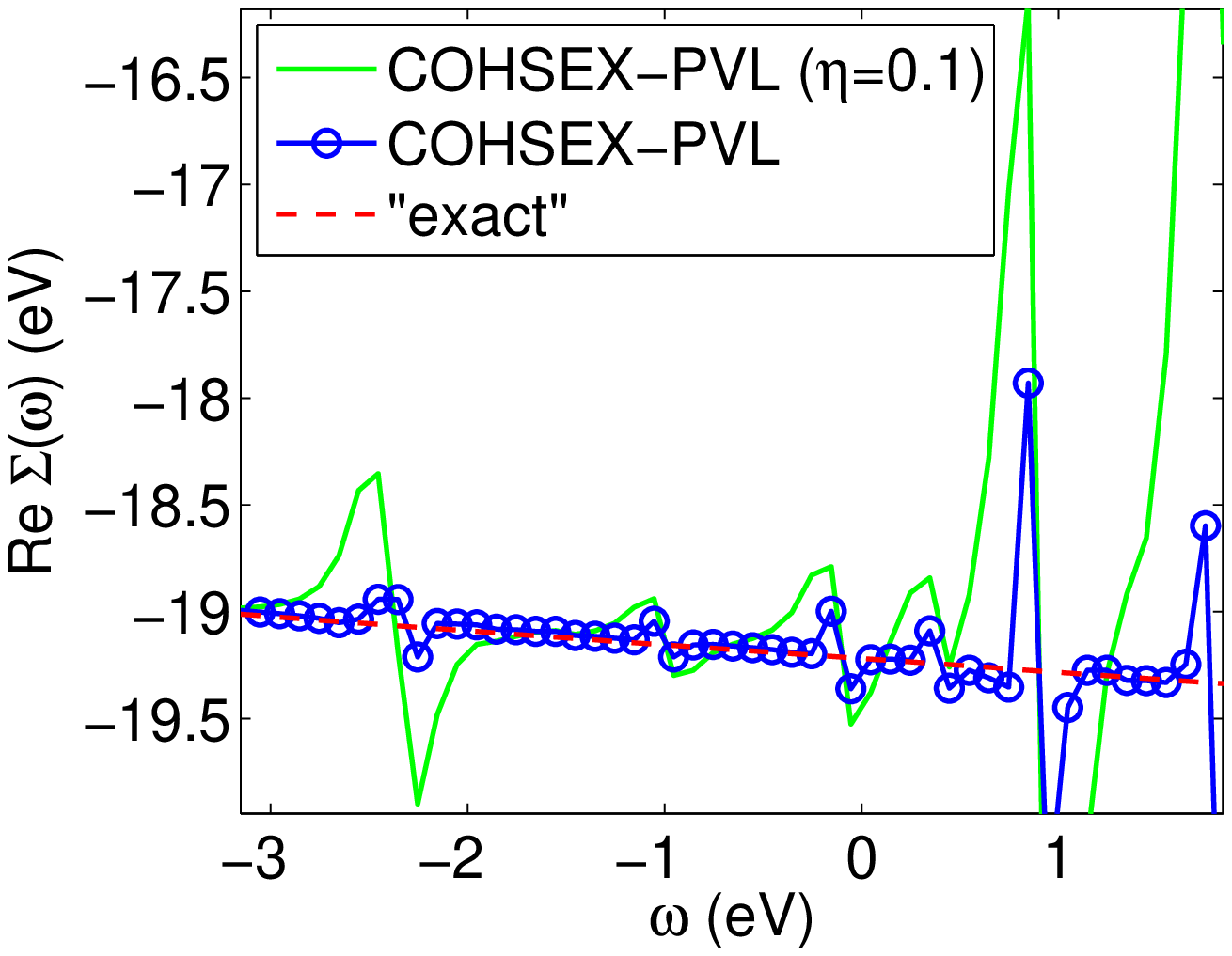}
\caption{The real part of $\langle \phi_4|\Sigma(\omega)|\phi_4 \rangle$ for a
methane molecule computed by using~\eqref{eq:splitlp} and~\eqref{eq:kresse}
to the XCOR (left) and the COHSEX (right) expressions.}\label{fig:etalimit}
\end{figure}

In Figure~\ref{fig:etalimit}, we demonstrate the benefit of
taking the $\eta \rightarrow 0$ limit in the denominator of~\eqref{eq:prototype} first before performing the principal value integration. Without
taking this limit first, the principal value integration based on
~\eqref{eq:kresse} produces much larger error for both the COHSEX
and XCOR expressions in different frequency regions.

\section{Conclusion}
We presented a technique for performing numerical convolution of
a Green's function with a screened Coulomb potential for the GW
approximation of the self energy term in a Dyson's equation.
Our numerical integration is performed directly on the real frequency axis.
To overcome the difficulty associated with the singularities of
the integrand, we take the zero broadening limit in Green's
function first and replace the numerator of the
integrand with piecewise polynomial approximations so that the
principal value integral of the approximate integrand can be obtained
analytically. We presented the error bound associated with this
integration scheme and showed by numerical examples that this
technique produced more accurate results than standard numerical
quadrature rules such as the trapezoidal rule. Consequently, to
achieve the same level accuracy, fewer quadrature points are
needed. This leads to a reduction in computational cost. We also showed
that applying the same numerical integration to different expressions
of the GW self energy approximation (e.g. the full-frequency COHSEX and XCOR) may
lead to different levels of numerical accuracy. For a given frequency,
one expression may be preferred over the other. Our technique
has been implemented in the BerkeleyGW software package \cite{Deslippe12},
and it gives a significant improvement over the previous integration
scheme.  We should mention that there are other techniques for
treating singularities of the integrand in the GW convolution \cite{Daling91,
Jin99,Rojas95,Kotani02,Lebegue03}.
We will compare the efficiency and accuracy of the technique presented
in this paper with other techniques in our future work. Moreover, in this work both
theoretical analysis and numerical results are for molecules. We
will generalize and apply our technique to the GW approximation for solids in the future.

\section*{Acknowledgements}
Partial support for this work was provided through Scientific Discovery through
Advanced Computing (SciDAC) program funded by U.S. Department of Energy,
Office of Science, Advanced Scientific Computing Research, Basic Energy Sciences,
and the U.S. Department of Energy under contract number DE-AC02-05CH11231.
The computational results were obtained at the National Energy Research Scientific
Computing Center (NERSC), which is supported by the Director, Office of Advanced Scientific
Computing Research of the U.S. Department of Energy under contract number DE-AC02-05CH11232.
Work at the Molecular Foundry was supported by the Office of Science,
Office of Basic Energy Sciences, of the U.S. Department of Energy under contract number DE-AC02-05CH11231.
F.L. is also grateful to the
support from the National Science Foundation of China (grants 11071265 and 11171232),
the China Scholarship Council, and Program for Innovation Research
in Central University of Finance and Economics.
This work was completed during her visit to Lawrence Berkeley National Laboratory.
L.L. and A.F.K also acknowledge the support by the Laboratory Directed Research and
Development Program of Lawrence Berkeley National Laboratory under
the U.S. Department of Energy contract number DE-AC02-05CH11231.
S.G.L. acknowledges the support of a Simons Foundation Fellowship in Theoretical Physics.

\bibliographystyle{unsrt}
\bibliography{sigma}

\end{document}